\documentclass[runningheads]{llncs}
\usepackage[T1]{fontenc}
\usepackage{lmodern}   
\usepackage{graphicx}
\usepackage{booktabs}
\usepackage{amsmath,amssymb}
\usepackage{array}
\usepackage{xcolor}
\usepackage{colortbl}
\usepackage{tikz}
\usetikzlibrary{positioning,calc,fit,arrows.meta,shapes.misc}
\usepackage{pifont}
\usepackage{float}
\usepackage{tabularx}
\usepackage[hidelinks]{hyperref}
\newcommand{\chip}[2]{{\setlength{\fboxsep}{1.6pt}\colorbox{#1!14}{\textcolor{#1!80!black}{\scriptsize #2}}}}
\newcommand{\ok}{\ding{51}}\newcommand{\bad}{\ding{55}}
\tikzset{
  pics/icon doc/.style={code={\draw[white, line width=0.7pt, line join=round] (0,0) -- (0,0.3) -- (0.15,0.3) -- (0.23,0.22) -- (0.23,0) -- cycle; \draw[white, line width=0.5pt] (0.05,0.07)--(0.18,0.07) (0.05,0.13)--(0.18,0.13) (0.05,0.19)--(0.12,0.19);}},
  pics/icon flask/.style={code={\draw[white, line width=0.7pt, line join=round] (0.075,0.3)--(0.075,0.17)--(0,0.02)--(0.23,0.02)--(0.155,0.17)--(0.155,0.3); \draw[white, line width=0.7pt] (0.04,0.3)--(0.19,0.3); \fill[white] (0.035,0.085) -- (0.195,0.085) -- (0.155,0.165) -- (0.075,0.165) -- cycle;}},
  pics/icon stamp/.style={code={\draw[white, line width=0.7pt] (0.115,0.15) circle (0.125); \draw[white, line width=0.9pt, line join=round] (0.06,0.15)--(0.10,0.10)--(0.18,0.21);}},
  pics/icon lock/.style={code={\fill[white, rounded corners=0.5pt] (0.02,0) rectangle (0.21,0.15); \draw[white, line width=0.7pt] (0.06,0.15) -- (0.06,0.22) arc (180:0:0.055) -- (0.17,0.15);}},
  pics/icon gate/.style={code={\draw[white, line width=0.7pt, line join=round] (0.115,0.32) -- (0.23,0.26) -- (0.21,0.1) -- (0.115,0) -- (0.02,0.1) -- (0,0.26) -- cycle; \draw[white, line width=0.9pt, line join=round] (0.06,0.15)--(0.10,0.10)--(0.17,0.21);}},
  pics/icon browser/.style={code={\draw[line width=0.6pt, rounded corners=0.8pt] (0,0.03) rectangle (0.3,0.27); \draw[line width=0.5pt] (0,0.2)--(0.3,0.2); \fill (0.045,0.235) circle (0.014) (0.085,0.235) circle (0.014) (0.125,0.235) circle (0.014); \draw[line width=0.5pt] (0.05,0.14)--(0.21,0.14) (0.05,0.085)--(0.16,0.085);}},
  pics/icon terminal/.style={code={\draw[line width=0.6pt, rounded corners=0.8pt] (0,0.03) rectangle (0.3,0.27); \draw[line width=0.7pt, line cap=round, line join=round] (0.055,0.205)--(0.115,0.15)--(0.055,0.095); \draw[line width=0.7pt, line cap=round] (0.145,0.09)--(0.24,0.09);}},
  pics/icon code/.style={code={\draw[line width=0.7pt, line cap=round, line join=round] (0.08,0.24)--(0.005,0.15)--(0.08,0.06) (0.22,0.24)--(0.295,0.15)--(0.22,0.06) (0.18,0.27)--(0.12,0.03);}},
  pics/icon gear/.style={code={\begin{scope}[shift={(0.15,0.15)}]\foreach \a in {0,45,...,315} {\fill[rotate=\a] (-0.032,0.08) rectangle (0.032,0.145);} \fill[even odd rule] (0,0) circle (0.108) (0,0) circle (0.045);\end{scope}}},
  pics/icon chip/.style={code={\draw[line width=0.6pt, rounded corners=0.6pt] (0.055,0.055) rectangle (0.245,0.245); \foreach \t in {0.1,0.15,0.2} {\draw[line width=0.45pt] (\t,0.245)--(\t,0.3) (\t,0.055)--(\t,0) (0.055,\t)--(0,\t) (0.245,\t)--(0.3,\t);} \draw[line width=0.6pt] (0.115,0.1)--(0.115,0.2) (0.115,0.15)--(0.195,0.15);}},
  pics/icon robot/.style={code={\draw[line width=0.6pt, rounded corners=1.2pt] (0.025,0.02) rectangle (0.275,0.205); \fill (0.1,0.125) circle (0.026) (0.2,0.125) circle (0.026); \draw[line width=0.5pt, line cap=round] (0.11,0.065)--(0.19,0.065); \draw[line width=0.5pt] (0.15,0.205)--(0.15,0.25); \fill (0.15,0.27) circle (0.026);}},
  pics/icon db/.style={code={\draw[line width=0.6pt] (0.15,0.25) ellipse (0.12 and 0.04); \draw[line width=0.6pt] (0.03,0.25)--(0.03,0.06) arc (180:360:0.12 and 0.04)--(0.27,0.25); \draw[line width=0.5pt] (0.03,0.155) arc (180:360:0.12 and 0.04);}},
  pics/icon branch/.style={code={\draw[line width=0.55pt] (0.07,0.04)--(0.07,0.26) (0.07,0.1) .. controls (0.07,0.19) and (0.21,0.13) .. (0.21,0.22); \fill (0.07,0.04) circle (0.035) (0.07,0.26) circle (0.035) (0.21,0.24) circle (0.035);}},
  pics/icon loop/.style={code={\draw[line width=0.65pt, -{Latex[length=2.4pt, width=3pt]}] (0.15,0.265) arc (90:385:0.115);}},
  pics/icon cloud/.style={code={\fill (0.09,0.115) circle (0.065) (0.165,0.17) circle (0.085) (0.235,0.115) circle (0.06); \fill[rounded corners=0.6pt] (0.025,0.05) rectangle (0.295,0.12);}},
  pics/icon check/.style={code={\draw[line width=0.8pt, line cap=round, line join=round] (0.03,0.15)--(0.11,0.06)--(0.27,0.25);}},
}
\def\cw{2.75}\def\ch{2.95}
\newcommand{\revcard}[7]{%
  \begin{scope}
    \clip[rounded corners=2.5pt] (#1,0) rectangle ({#1+\cw},{-\ch});
    \fill[white] (#1,0) rectangle ({#1+\cw},{-\ch});
    \fill[#2] (#1,0) rectangle ({#1+\cw},-0.44);
  \end{scope}
  \draw[#2!60, rounded corners=2.5pt] (#1,0) rectangle ({#1+\cw},{-\ch});
  \pic at ({#1+0.12},-0.38) {icon #3};
  \node[anchor=west, text=white, font=\scriptsize\bfseries, inner sep=0] at ({#1+0.48},-0.22) {#4};
  \node[anchor=north west, inner sep=0] at ({#1+0.11},-0.58) {\parbox[t][2.25cm][t]{2.53cm}{\raggedright\scriptsize #7}};
  \node[inner sep=0, font=\scriptsize] at ({#1+\cw/2},0.55) {#5};
  \node[inner sep=0, font=\scriptsize] at ({#1+\cw/2},{-\ch-0.77}) {#6};
}
\newcommand{\archcard}[9]{%
  \coordinate (#1-nw) at (#2,#3); \coordinate (#1-se) at ({#2+#4},{#3-#5});
  \begin{scope}
    \clip[rounded corners=2.5pt] (#1-nw) rectangle (#1-se);
    \fill[white] (#1-nw) rectangle (#1-se);
    \fill[#6] (#1-nw) rectangle ({#2+#4},{#3-0.44});
  \end{scope}
  \draw[#6!60, rounded corners=2.5pt] (#1-nw) rectangle (#1-se);
  \pic[white] at ({#2+0.1},{#3-0.37}) {icon #7};
  \node[anchor=west, text=white, font=\scriptsize\bfseries, inner sep=0] at ({#2+0.5},{#3-0.22}) {#8};
  \node[anchor=east, text=white, font=\scriptsize\itshape, inner sep=0] at ({#2+#4-0.12},{#3-0.22}) {#9};
}
\newcommand{\archtile}[6]{%
  \fill[#3, rounded corners=1.5pt] (#1,#2) rectangle ({#1+0.44},{#2-0.44});
  \pic[white] at ({#1+#5},{#2-0.37}) {icon #4};
  \node[anchor=west, align=left, inner sep=0, font=\scriptsize] at ({#1+0.55},{#2-0.22}) {\strut#6\strut};
}
\definecolor{otblue}{HTML}{0072B2}
\definecolor{otorange}{HTML}{E69F00}
\definecolor{otteal}{HTML}{009E73}
\definecolor{otpurple}{HTML}{CC79A7}
\definecolor{otgray}{HTML}{5A5A5A}
\definecolor{otred}{HTML}{D55E00}

\newcommand{\pp}{ProofPatch}
\newcommand{\lean}[1]{\texttt{#1}}

\begin{document}
\widowpenalty=300 \clubpenalty=300
\title{\pp: Reviewing Lean Statements and Keeping the Reviews Valid as Formalizations Evolve}
\titlerunning{\pp}
\author{Xujun Che \and Depeng Xu \and Jian Xiang \and Taufiqua Khan}
\authorrunning{X. Che et al.}
\institute{University of North Carolina at Charlotte, Charlotte, NC, USA\\
\email{\char123 xche,dxu7,jxiang1,tkhan13\char125 @charlotte.edu}}
\maketitle

\begin{abstract}
The kernel of the Lean proof assistant checks that a proof establishes its statement. Whether the statement says what was intended is left to human reviewers. This is the old problem of validating a specification, and it grows more pressing as language models write formal statements in bulk and agents edit formal developments continuously. We present \pp{}, a tool that helps reviewers judge Lean statements and keeps their judgments valid while the development changes. A reviewer can ask the kernel targeted questions about a statement, such as whether a hypothesis is needed or whether a counterexample exists. Each decision is stored with fingerprints of the statement as Lean elaborated it and of the definitions it depends on, and any declaration can be pinned so that a patch changing it is blocked, whether a person or an agent built on a large language model (LLM) wrote the patch. Working on elaborated terms rather than source text is the central design choice: notation and formatting changes leave a decision intact, and a changed definition reaches every theorem that uses it. On 229 pairs of consecutive revisions of Mathlib, Lean's mathematics library, \pp{} sends 118 of 12,898 synthetic decisions back for review. Every text-based alternative misses at least 38 of these or sends back at least 13 times as many. On 1,527 labeled statements, kernel evidence flags unfaithful ones with precision 0.88 to 0.96, and in 344 runs of an LLM agent no change to a pinned statement got through.
\end{abstract}

\section{Introduction}
\label{sec:intro}

A machine-checked proof is only as good as its statement. The kernel of the Lean proof assistant~\cite{demoura2021lean4}, its small trusted proof checker, guarantees that a proof establishes the statement it is given, but not that the statement is the theorem anyone meant. Checking that is the long-standing problem of validating a specification~\cite{pollack1998believe,woodcock2009fm}, and in practice it falls to people who read the formal statement next to its informal counterpart. Tools help them: counterexample generators in Isabelle~\cite{blanchette2010nitpick,bulwahn2012quickcheck} and Coq~\cite{paraskevopoulou2015quickchick} refute wrong conjectures before anyone tries a proof, and vacuity detection~\cite{beer2001vacuity,kupferman2003vacuity} catches properties that hold for a trivial reason.

Two developments make this review harder. The first is volume. Large language models (LLMs) now prove competition problems in Lean~\cite{ren2025deepseekproverv2} and translate informal statements into Lean at scale~\cite{weng2025survey}, and compiling is a weak sign of being right: on 227 graduate-level statements, a tool-augmented GPT-5.2 agent compiles 87.2\,\% but is faithful on only 57.7\,\%~\cite{zhang2026beyond}. The second is durability. Formal developments change constantly, increasingly through agents, and a review decision goes stale without notice when something it relied on changes. A theorem can keep every character of its statement while a definition it uses is rewritten. A line diff then shows nothing on the theorem, and continuous integration re-checks the proofs but not the review.

We present \pp{}, a tool for reviewing Lean statements and keeping the outcome of that review valid as the development evolves. Its central design choice is to work on \emph{elaborated terms} rather than on source text. An elaborated term is the statement as the kernel sees it, after Lean has expanded notation, inserted implicit arguments and coercions, and resolved type-class instances. Two commits to Mathlib, Lean's community library of mathematics~\cite{mathlib2020}, show why this matters. One rewrote statements such as \mbox{\lean{ContMDiffWithinAt I I\textquotesingle{} n f s x}} into a new notation, \mbox{\lean{CMDiffAt[s] n f x}}. The text changed, but the elaborated statement stayed the same up to \emph{definitional equality}, the equality the kernel itself checks by unfolding definitions and computing, so a review of the old statement still applies. Another replaced \mbox{\lean{open Classical in}} by \mbox{\lean{open scoped Classical in}} above a definition. The definition's text stayed the same, but its elaborated body now uses a different decidability instance, the term that decides the condition of an if-then-else. Reviews of the 26 theorems whose statements use the definition relied on the old body, yet a line diff shows nothing on any of them.

\begin{figure}[t]
\centering
\begin{tikzpicture}[arr/.style={-{Latex[length=4pt]}, otgray, line width=0.8pt}]
\fill[otgray!7] (0,0.28) rectangle (12.2,0.82);
\fill[otgray!7] (0,{-\ch-0.5}) rectangle (12.2,{-\ch-1.04});
\node[anchor=south west, fill=otgray!18, text=otgray!80!black, font=\scriptsize\itshape, inner sep=2pt, rounded corners=1pt] at (0,0.82) {reviewer};
\node[anchor=south west, fill=otgray!18, text=otgray!80!black, font=\scriptsize\itshape, inner sep=2pt, rounded corners=1pt] at (0,{-\ch-0.5}) {kernel};
\revcard{0}{otblue}{doc}{read}{reads both statements}{elaborates}{%
  \textbf{informal}\\ non-empty interval:\\ the clamped value\\ never exceeds \lean{hi}\\[2pt]
  \textbf{formal}\\ \lean{(h : lo ${\le}$ hi) :}\\ \lean{clamp lo hi x ${\le}$ hi}}
\revcard{3.15}{otorange}{flask}{question}{drops a hypothesis}{replays, proves, refutes}{%
  \textbf{challenge} drop \lean{h}\\[2pt]
  proof \chip{otred}{\bad\ fails}\\[2pt]
  counterexample\\ \lean{lo=1 hi=0 x=1}\\[2pt]
  \chip{otorange}{h is needed}}
\revcard{6.30}{otgray}{stamp}{record}{accepts and pins}{computes fingerprints}{%
  \textbf{receipt} by alice\\ cites the challenge\\[2pt]
  \textbf{fingerprints}\\ statement, \lean{clamp},\\ environment\\[2pt]
  \textbf{locks} statement,\\ body of \lean{clamp}}
\revcard{9.45}{otteal}{gate}{re-check}{reads the verdict}{compares}{%
  \textbf{patch}\\ rewrites \lean{clamp}\\[2pt]
  \chip{otred}{\bad\ lock on \lean{clamp}}\\[2pt]
  6 receipts sent back\\ 1 receipt kept\\[2pt]
  \chip{otred}{blocked}}
\foreach \x in {2.75,5.90,9.05} {\draw[arr] ({\x+0.05},-1.5) -- ({\x+0.35},-1.5);}
\end{tikzpicture}
\caption{One review in \pp{}. The reviewer acts on the artifacts in the middle row. Every fact in them comes from the kernel and from elaborated terms, not from text or from a model.}
\label{fig:loop}
\end{figure}
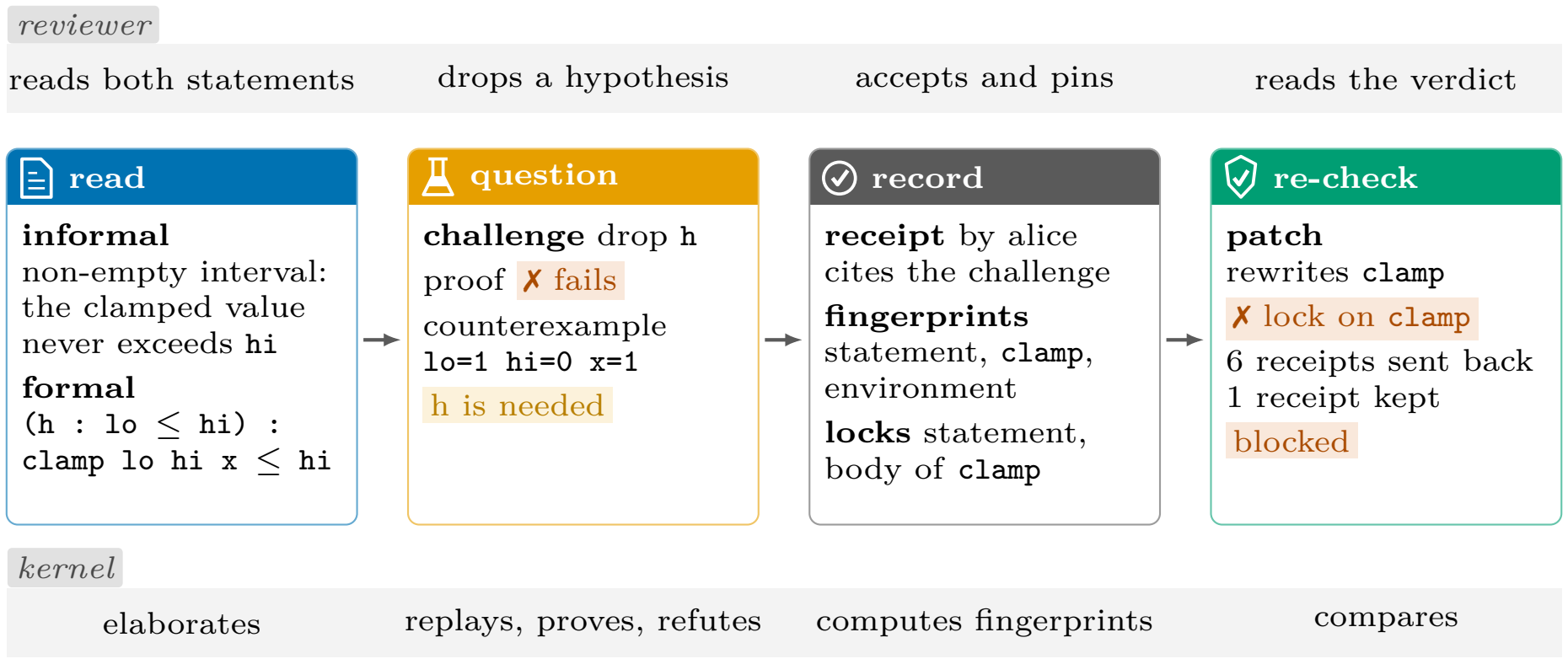

On this choice \pp{} builds the review loop of Figure~\ref{fig:loop}. A reviewer can \emph{challenge} a statement: modify it in a small, typed way, for instance by dropping a hypothesis, and let the kernel check the result, which yields evidence such as a counterexample or a proof that the hypothesis was never needed. The decision is recorded as a \emph{receipt}, which stores fingerprints of the elaborated statement and of every definition it depends on, so that it can be re-checked mechanically after any change. A reviewer can also \emph{lock} a declaration to pin it. A \emph{gate} runs on every patch, written by a person or by an LLM agent, and reports which receipts no longer apply and which locks are violated. Two versions of a declaration count as the same when their elaborated terms coincide after a small canonicalization, or when the kernel confirms that they are definitionally equal. This notion is precise and checkable, and deliberately conservative: a change the kernel cannot see through goes back to the reviewer.

\pp{} applies to any Lean development. Machine-generated formalizations make it necessary by multiplying what there is to review and how often it changes. This paper makes four contributions:
\begin{itemize}
\item a way to record review decisions against elaborated statements and their dependencies, with precise guarantees about what a kept receipt implies;
\item kernel-based evidence for reviewers, obtained by counterexample search and mutation on elaborated statements, with the guarantee that the statement the kernel checks is the statement the reviewer sees;
\item a gate that applies these checks to every patch, in continuous integration, in the editor and inside an agent loop that feeds the gate's report back to the model, implemented for Lean~4 and Mathlib;
\item an evaluation on 229 pairs of Mathlib revisions against text-based baselines, on 1,527 labeled statements, and on 344 agent runs.
\end{itemize}

\section{\pp{} at a glance}
\label{sec:example}

This section follows one review through the tool, using the four notions introduced informally above: challenges, receipts, locks and the gate. The next three sections make them precise.

\subsection{The example and its review state}

Figure~\ref{fig:example} lists the running example. The definition \lean{clamp lo hi x} moves an integer \lean{x} into the interval from \lean{lo} to \lean{hi}, and \lean{score x} clamps it into $[0,100]$. Five theorems state properties of the two. The remaining four illustrate further kinds of evidence. Every theorem carries an informal statement, which the tool shows next to the formal one. For \lean{clamp\_le\_hi} it reads \emph{for any non-empty interval ($\mathit{lo} \le \mathit{hi}$), the clamped value never exceeds $\mathit{hi}$}.

\begin{figure}[t]
\centering
\scriptsize
\setlength{\tabcolsep}{2.6pt}\renewcommand{\arraystretch}{1.06}
\newcommand{\hit}{\rowcolor{otred!7}}
\newcommand{\acc}[1]{\chip{otblue}{\ok\,#1}}
\newcommand{\rej}[1]{\chip{otred}{\bad\,#1}}
\newcommand{\back}{\chip{otorange}{re-review}}
\begin{tabular}{@{}lllc@{}}
\toprule
declaration, proof elided & receipts & lock & after patch \\
\midrule
\hit \lean{def clamp (lo hi x : $\mathbb{Z}$) : $\mathbb{Z}$ :=} & & definition & \chip{otred}{\bad\,lock fails} \\
\hit \lean{\ \ \ \ max lo (min hi x)} & & & \\
\hit \lean{theorem clamp\_le\_hi (lo hi x : $\mathbb{Z}$) (h : lo $\le$ hi) :} & \acc{alice} & statement & \back \\
\hit \lean{\ \ \ \ clamp lo hi x $\le$ hi} & & & \\
\hit \lean{theorem lo\_le\_clamp (lo hi x : $\mathbb{Z}$) (h : lo $\le$ hi) :} & \rej{alice} & & \back \\
\hit \lean{\ \ \ \ lo $\le$ clamp lo hi x} & & & \\
\hit \lean{theorem clamp\_of\_mem (lo hi x : $\mathbb{Z}$) (h$_1$ : lo $\le$ x)} & \acc{alice} & statement & \back \\
\hit \lean{\ \ \ \ (h$_2$ : x $\le$ hi) : clamp lo hi x = x} & & & \\
\lean{theorem two\_mul\_le\_sq\_add\_sq (a b : $\mathbb{Z}$) :} & \acc{alice} & statement & \chip{otteal}{\ok\,kept} \\
\lean{\ \ \ \ 2 * a * b $\le$ a * a + b * b} & & & \\
\hit \lean{def score (x : $\mathbb{Z}$) : $\mathbb{Z}$ := clamp 0 100 x} & & & \\
\hit \lean{theorem score\_nonneg (x : $\mathbb{Z}$) : 0 $\le$ score x} & \acc{alice} \acc{bob} & statement & \chip{otorange}{re-review $\times$2} \\
\hit \lean{theorem score\_le\_hundred (x : $\mathbb{Z}$) :} & \acc{bob} & statement & \back \\
\hit \lean{\ \ \ \ score x $\le$ 100} & & & \\
\lean{theorem sub\_add\_cancel\_of\_le (n m : $\mathbb{N}$) (h : m $\le$ n) :} & & & \\
\lean{\ \ \ \ n - m + m = n} & & & \\
\lean{theorem vacuous (n : $\mathbb{N}$) (h$_1$ : n < 3) (h$_2$ : 5 < n) :} & & & \\
\lean{\ \ \ \ n = 100} & & & \\
\lean{theorem strict\_of\_le (n : $\mathbb{N}$) (h : 0 $\le$ n) :} & & & \\
\lean{\ \ \ \ n $\le$ n \textasciicircum{} 2} & & & \\
\bottomrule
\end{tabular}
\caption{The running example with its review state, and the gate's verdict on a patch that rewrites the body of \lean{clamp} as \lean{min hi (max lo x)}. Shaded: \lean{clamp} and the declarations whose statements depend on it. There are seven receipts on six theorems, five statement locks and one definition lock.}
\label{fig:example}
\end{figure}

\subsection{Reviewing a statement}

\noindent\begin{minipage}{\textwidth}
A reviewer opens \lean{clamp\_le\_hi}:
\begin{quote}
\lean{theorem clamp\_le\_hi (lo hi x : $\mathbb{Z}$) (h : lo $\le$ hi) :}\\
\hspace*{2em}\lean{clamp lo hi x $\le$ hi}
\end{quote}
\end{minipage}
The kernel has already checked the proof. The reviewer must decide whether the formal statement says the same as the informal one, and the hypothesis \lean{h} is the obvious place to look: does it express the non-empty interval, and is it needed? The tool shows the statement split into its parts, namely the variables, the hypothesis and the conclusion, each with the challenges that apply to it. Dropping a hypothesis is one of them. Others substitute a boundary value, change a constant or strengthen the conclusion.

When the reviewer drops \lean{h}, the tool elaborates the modified statement in an isolated branch of the file, replays the original proof, tries a fixed set of automatic provers and searches for a counterexample. The proof fails, and random testing finds \lean{lo = 1}, \lean{hi = 0}, \lean{x = 1}, for which \lean{clamp} returns 1, above the upper bound 0. The hypothesis is needed: an empty interval has no meaningful clamp. The reviewer, alice, accepts the theorem. The receipt cites this evidence and stores fingerprints of the statement and of \lean{clamp}, the definition it depends on. By project policy, accepting a theorem also locks its statement.

The same challenge on \lean{lo\_le\_clamp} tells a different story. The original proof still goes through without \lean{h}, so the hypothesis was never used, and the informal statement has no such assumption. Here alice rejects the theorem because of the spurious hypothesis, and also locks the body of \lean{clamp}, since every statement about \lean{clamp} depends on it. Figure~\ref{fig:example} shows the resulting review state: seven receipts on six theorems, two of them on \lean{score\_nonneg}, five statement locks and one definition lock.

\subsection{A patch}

Later, an agent or a colleague rewrites the body of \lean{clamp} as \mbox{\lean{min hi (max lo x)}}. The two bodies agree whenever \lean{lo $\le$ hi} and differ otherwise: \mbox{\lean{clamp 5 0 7}} was 5 and is now 0. Every proof in the file still compiles, and a line diff shows one changed line, inside a definition that has no receipt. The gate blocks the patch because the lock on the body of \lean{clamp} fails. It also sends six of the seven receipts back for re-review: those on the five theorems whose statements depend on \lean{clamp}, directly or through \lean{score}. Two of these, \lean{score\_nonneg} and \lean{score\_le\_hundred}, never mention \lean{clamp} in their text. The receipt on \lean{two\_mul\_le\_sq\_add\_sq}, which does not depend on \lean{clamp}, is kept, and the five statement locks hold, because no statement changed.

The reviewer now decides whether the six decisions still stand. They do: all five theorems still hold, and \lean{score} is the same function as before, because its interval is not empty. What changed is \lean{clamp} on empty intervals, and that is what the definition lock reports.

\section{Receipts and locks}
\label{sec:contracts}

This section defines what a receipt records and when it stays valid. The next two describe how the evidence it cites is produced and how the gate applies receipts and locks to a patch.

\subsection{Elaborated statements and fingerprints}

Lean elaborates a declaration's source into a term of its type theory, expanding notation, filling in implicit arguments, resolving type-class instances and inserting coercions. Two sources can elaborate to the same term, and one source can elaborate to different terms in different contexts~\cite{wiedijk2012pollack}, so \pp{} fingerprints elaborated terms. The \emph{statement fingerprint} $F_T(d)$ of a declaration $d$ is a hash of a canonical serialization of its \emph{elaborated statement}, that is, its elaborated type. A definition also has a \emph{value fingerprint} $F_V(d)$ of its body. Proofs are not fingerprinted: the kernel checks them, and they are not what a reviewer decides about.

Canonicalization removes differences that do not affect the logical content of a term: it erases binder names, renames universe parameters by position and drops metadata. It also applies three normalizations that real Mathlib commits showed to be necessary. It unfolds identity wrappers such as \lean{id} and \lean{inferInstanceAs}. It reduces a type-class argument reached through a chain of projections to the instance it denotes, because the chain the elaborator picks depends on the imports. Finally, it erases the source positions that \lean{panic!} writes into its error message, which the logical definition of \lean{panic} ignores. Each step replaces a term by a definitionally equal one, so equal fingerprints imply definitionally equal terms, up to hash collisions. Binder kinds, such as explicit or instance-implicit, are kept as part of a declaration's interface.

\subsection{Statement dependencies and closure fingerprints}
\label{sec:closure}

A declaration depends on other declarations in two ways. Its \emph{statement dependencies} are the constants in its elaborated statement and, for a definition, in its body. A change to one of them can change what the declaration says. Its \emph{proof dependencies} are the constants that occur only in a theorem's proof. A change to one of them can break the proof, which the kernel reports, but cannot change the statement. Review follows statement dependencies only, and the declarations reachable from $d$ through them form its \emph{dependency closure}. In Figure~\ref{fig:example}, \lean{score\_nonneg} depends on \lean{score}, which depends on \lean{clamp}, while its proof uses Mathlib lemmas that are irrelevant to its review.

The \emph{closure fingerprint} $R(d)$ summarizes a declaration together with everything its statement depends on:
\[ R(d) = H\bigl(F(d),\ \{(u, R(u)) : u \in D(d)\}\bigr), \]
where $H$ is a cryptographic hash function, $D(d)$ is the set of statement dependencies of $d$ inside the project, and $F(d)$ is $F_T(d)$ for a theorem and the pair $(F_T(d), F_V(d))$ for a definition. Mutually recursive declarations form a single node of the dependency graph, whose members enter with their names. Declarations outside the project, from Mathlib or the core library, enter by name only, because the environment identity below pins their content. The name of $d$ itself enters neither $F(d)$ nor $R(d)$, so renaming a declaration leaves its closure fingerprint unchanged. A dependent whose statement mentions it does change, because its elaborated statement now refers to the new name.

\subsection{Environment identity}

Fingerprints are comparable only within one environment. The \emph{environment identity} $E$ hashes the Lean toolchain, the dependency manifest, which pins every package such as Mathlib to a commit, and the source of the project's own modules that the file imports. Adding or removing an import of a pinned package does not change $E$, because every constant that an elaborated term mentions is already pinned. Changing the toolchain, the manifest or an imported project module does change $E$. Every receipt on the file is then sent back for review, and every lock on it fails.

\subsection{Validity of receipts and locks}

A \emph{receipt} $\rho = (r, d, \delta, h_\iota, F, R, E)$ records that reviewer $r$ made decision $\delta$, one of accept, reject or unsure, about declaration $d$, against the informal statement with hash $h_\iota$, at a time when the declaration had fingerprint $F$, closure fingerprint $R$ and environment $E$. In a later revision $P'$ of the project $P$, the receipt is \emph{kept} if the informal statement, $F$, $R$ and $E$ are all unchanged, up to the definitional rewrites described below. Otherwise it is \emph{invalidated} and sent back to the reviewer with the first reason that applies: the informal statement changed, the environment changed, the declaration is gone, its statement changed, or a declaration in its closure changed.

A \emph{lock} pins one declaration or the whole project. A statement lock on $d$ holds in $P'$ if $d$ still exists there with the same $E$ and $F_T$, again up to definitional rewrites. A definition lock also requires the same $F_V$. A trust lock holds if no declaration in its scope depends on an axiom outside an allowed set. By default this set contains Lean's three standard axioms, \lean{propext}, \lean{Classical.choice} and \lean{Quot.sound}, and excludes \lean{sorryAx}, the axiom behind \lean{sorry}, Lean's placeholder for a missing proof. An environment lock holds if $E$ is unchanged. A lock whose target has disappeared fails.

\subsection{What the gate guarantees}
\label{sec:guarantees}

Write $t \equiv t'$ when two elaborated terms have equal fingerprints, or when the kernel has confirmed that they are definitionally equal. In both cases $t$ and $t'$ are definitionally equal, so $\equiv$ is a sound approximation of definitional equality.

\begin{proposition}[Locks]\label{prop:locks}
If the gate accepts a revision $P'$ of $P$, then for every statement lock on $d$ the elaborated statement of $d$ in $P'$ is $\equiv$ to the locked one, so the kernel accepts a proof of either as a proof of the other. For every definition lock, the same holds for the body. Every declaration in the scope of a trust lock depends only on allowed axioms.
\end{proposition}
\emph{Proof sketch.} The gate accepts only if every lock holds, and a statement lock holds only if $E$ is unchanged and either $F_T$ is unchanged or the kernel confirmed definitional equality in the new environment. Both imply, up to hash collisions, that the two statements are definitionally equal, and the kernel's conversion rule then lets a proof of one stand for the other. Definition locks add the same argument for $F_V$. Trust locks are checked against the axioms the kernel reports for each declaration of $P'$. \hfill$\square$

\begin{proposition}[Receipts]\label{prop:receipts}
If a receipt on $d$ is kept from $P$ to $P'$, then the informal statement and the environment are unchanged, and the elaborated statement of $d$, as well as the statement and body of every project declaration in its dependency closure, is $\equiv$ to the version that was reviewed.
\end{proposition}
\emph{Proof sketch.} A receipt is kept only if $h_\iota$, $E$, $F(d)$ and $R(d)$ are unchanged, where $R$ uses the old fingerprint for every declaration the kernel confirmed definitionally equal. Since $R$ hashes $F$ with the closure fingerprints of the named dependencies, recursively, an unchanged $R$ implies, up to hash collisions, unchanged fingerprints or confirmed definitional equality along the whole closure. Declarations outside the project are pinned by $E$. \hfill$\square$

Invalidation is safe in the sense of regression test selection~\cite{rothermel1997safe}: whenever a term in the closure changes beyond $\equiv$, the receipt goes back to the reviewer. This covers definitions that are provably but not definitionally equal, such as the one in the introduction, and definitional equalities that the kernel's incomplete check misses~\cite{carneiro2019lean}. The reviewer then decides again. Whether a statement is faithful in the first place is the reviewer's judgment, and the evidence that challenges produce supports it.

\subsection{Definitional rewrites and renames}
\label{sec:defeq}

Equal fingerprints are a stricter requirement than definitional equality. A Mathlib commit that removed \lean{set\_option backward.inferInstanceAs.wrap false} added an auxiliary definition to the elaborated body of an instance, and another commit spelled out a default field of a structure instance. The gate therefore has a second stage. When a declaration's fingerprint changed and every project declaration its old term depends on has an unchanged closure fingerprint, so that the old term can be read in the new environment, the gate asks the kernel whether the old and the new term are definitionally equal there, within a fixed resource budget. If the kernel confirms it, the change is reported as a definitional rewrite: locks hold, receipts are kept, and closure fingerprints are computed with the old fingerprint. A timeout counts as a change.

Renames are detected by fingerprint: a declaration that disappears under one name and reappears under another with the same kind, universe parameters and fingerprints is reported as renamed if the match is one-to-one. Its locks follow it, and its receipts are evaluated under the new name. Statements that mention it are reported as changed, because they refer to it by name. A rename combined with a change of statement counts as a removal and an addition.

\section{Challenges: evidence about a statement}
\label{sec:challenges}

Challenges turn the kernel into a source of evidence about a statement. A challenge mutates the statement, as in mutation testing~\cite{jia2011mutation}. A vacuity check applies vacuity detection~\cite{beer2001vacuity,kupferman2003vacuity} to theorem statements, and refutation combines decision procedures with random testing~\cite{claessen2000quickcheck}, as counterexample generators for proof assistants do~\cite{blanchette2010nitpick,bulwahn2012quickcheck,paraskevopoulou2015quickchick}. In \pp{} the operations act on elaborated statements, every outcome is reported with a fixed wording, and the statement the kernel checks is guaranteed to be the one the reviewer sees.

\subsection{Operations}

The tool presents a statement as its parts: the variables and hypotheses in front of the colon, and the conclusion. A hypothesis can be dropped, or weakened by turning $<$ into $\le$ or keeping one side of a conjunction. The conclusion can be strengthened by turning $\le$ into $<$ or keeping one side of a disjunction. The $k$-th occurrence of a numeral can be replaced, a variable can be instantiated with a term the reviewer types, and a definition can be replaced by an approved alternative. Each operation is a typed transformation of the elaborated statement, and one that would leave a dangling reference, such as dropping a hypothesis that another one mentions, is refused. Two further operations act on the whole statement: \emph{identity} searches for a counterexample to the statement as written, and \emph{vacuity} replaces the conclusion by \lean{False}, so that a proof of the result shows the hypotheses to be contradictory. Identity, vacuity and dropping each hypothesis need no input from the reviewer, so the tool runs them on every statement in advance. We call these challenges \emph{probes}.

\subsection{Judging a challenge}

The modified statement is elaborated in an isolated branch of the file, and four things are tried in order, as in the evidence of Figure~\ref{fig:card}. The original proof is replayed. Nine automatic provers are run, among them the simplifier \lean{simp}, the arithmetic procedures \lean{omega} and \lean{nlinarith}, the proof search \lean{aesop}~\cite{limperg2023aesop} and the decision procedure \lean{decide}. The negation is attempted with \lean{decide}, and random testing searches for a counterexample. The outcome is one of six grades, each with a fixed wording. \emph{Replay passes} means the original proof never used what was changed, and \emph{alternative proof} that the modified statement holds. \emph{Refuted} rests on a kernel-checked negation, \emph{counterexample candidate} on random testing. \emph{Proof sensitivity} means only the replay fails, and its wording says this does not make the assumption mathematically necessary. Otherwise the grade is \emph{inconclusive}.

\begin{figure}[t]
\centering
\scriptsize
\setlength{\tabcolsep}{3pt}\renewcommand{\arraystretch}{1.05}
\newcommand{\cardrow}[3]{\multicolumn{1}{@{}!{\color{#1}\vrule width 2.4pt}@{\hspace{4pt}}>{\raggedright\arraybackslash}p{#3}@{}}{#2}\\[1pt]}
\begin{tabular}{@{}>{\raggedright\arraybackslash}p{11.6cm}@{}}
\cellcolor{otgray}\textcolor{white}{\tikz[baseline=-2pt]\pic at (0,-0.2) {icon flask};\hspace{9pt}\textbf{challenge}\quad \lean{sub\_add\_cancel\_of\_le}\quad drop hypothesis \lean{h : m $\le$ n}}\\[2pt]
\cardrow{otblue}{\textbf{before}\ \ \lean{$\forall$ (n m : $\mathbb{N}$), m $\le$ n $\to$ n - m + m = n}\qquad \textbf{after}\ \ \lean{$\forall$ (n m : $\mathbb{N}$), n - m + m = n}}{11.2cm}
\cardrow{otorange}{\textbf{replay}\ \ \chip{otred}{\bad\ fails}\ \ original proof (\lean{omega}) on the new statement}{11.2cm}
\cardrow{otorange}{\textbf{alternatives}\ \ \chip{otred}{\bad\ none}\ \ none of the nine provers, among them \lean{simp}, \lean{omega}, \lean{nlinarith}, \lean{aesop}, \lean{decide}}{11.2cm}
\cardrow{otorange}{\textbf{refutation}\ \ \chip{otorange}{candidate}\ \ random testing: \lean{n = 0, m = 1} gives \lean{0 - 1 + 1 = 1 $\ne$ 0}. The negation is not kernel-checked}{11.2cm}
\cardrow{otorange}{\textbf{downstream}\ \ \chip{otgray}{none}\ \ no declaration depends on the changed statement}{11.2cm}
\cardrow{otteal}{\textbf{fidelity}\ \ \chip{otteal}{\ok\ executed = displayed}\ \ checked and displayed statement have one fingerprint}{11.2cm}
\cardrow{otgray}{\textbf{grade}\ \ \chip{otorange}{counterexample candidate}\ \ Random testing found a candidate counterexample to the modified statement. It is not kernel-checked.}{11.2cm}
\end{tabular}
\caption{The evidence of one challenge as the reviewer sees it, from a real run that took 0.4\,s. The hypothesis is needed, because subtraction on $\mathbb{N}$ truncates at zero. The last two rows are what a reviewer can rely on without reading the rest.}
\label{fig:card}
\end{figure}

\subsection{Fidelity}

The reviewer acts on a displayed statement and the kernel checks an elaborated one. \pp{} guarantees that they are the same: the branch records the fingerprint of the statement that Lean checked, the tool compares it with the fingerprint of the statement it displayed, and a mismatch is reported as an error without a grade. Every challenge is stored with both fingerprints, and a receipt cites the challenges it relies on, so its evidence can be re-run later.

\section{The gate}
\label{sec:gate}

\subsection{Verdict}

The gate compares two revisions of a file: the base from version control, and the head from the working tree, a commit or an agent's worktree. It classifies the effect of the patch on every declaration (added, removed, renamed, proof changed, statement changed, body changed, closure changed, definitional rewrite) and evaluates every lock and every receipt. The patch is accepted if the kernel accepts the head revision, every lock holds, no declaration gained a disallowed axiom, and the project's policy allows the number of invalidated receipts. Otherwise the report lists each violated lock and invalidated receipt with its reason. The same function drives continuous integration through its exit status, the browser interface, the editor and the agent loop.

\subsection{Policy and trust}

A project's policy fixes the allowed axioms, whether invalidated receipts block a patch, whether accepting a theorem locks it, which automatic provers challenges may use, and whether the definitional-equality stage runs. The trust check is per declaration, so a theorem that newly depends on \lean{sorryAx} is caught even if another theorem already did.

\subsection{Agents in the loop}

The agent loop is where machine-generated code enters \pp{}. An agent task names a file, a goal in natural language and an adapter, and runs in its own git worktree so that nothing it does touches the reviewer's files. The bundled adapter for language models sends the model the file, the goal and the contracts in force, that is, which statements are locked and which axioms are allowed, asks for the complete revised file, and runs the gate on it. If the gate blocks, the model receives the gate's report as feedback, and the loop repeats for a configured number of rounds. An accepted task is applied to the reviewer's files with one action, which also carries locks and receipts over renames. The gate, not the prompt, enforces the contracts: a model that is told not to change a locked statement and changes it anyway is refused like a person who does.

\section{Implementation}
\label{sec:impl}

All formal facts come from one long-running Lean~4 process per project, the \emph{kernel service}, driven over a JSON-lines protocol. It elaborates a file command by command and keeps the state after each command, so a challenge runs in the state just before its target, in a branch that never touches the file, and a reload after an edit re-elaborates only from the first changed command. Indexing yields, for every declaration, its kind, source range, fingerprints, statement and proof dependencies and axioms, and marks declarations that Lean generates automatically, such as recursors and equation lemmas, which reviewers do not see.

A daemon written in Python stores the review state as append-only JSON lines under version control, runs the gate and the agents, and serves the browser interface. A command-line tool exposes the same operations to continuous integration, and a VS Code extension shows each theorem's review status in the editor. Figure~\ref{fig:arch} in Appendix~\ref{app:demo} shows how these components fit together.

\section{Evaluation}
\label{sec:eval}

We ask five research questions (RQs). RQ1: do receipts follow real changes, and how do they compare with text-based methods? RQ2: can a reviewer rely on the evidence that challenges produce? RQ3: what does this evidence add to an LLM judge? RQ4: does the gate hold an LLM agent to its contracts? RQ5: is the tool fast enough?

\subsection{Data}

\emph{Revision pairs.} We took pairs of consecutive commits to one Mathlib~\cite{mathlib2020} file, from a ten-week window before a pinned Mathlib revision, and kept the 229 pairs in which both versions elaborate against that revision: the 40 pairs with the smallest diffs, all single-line, and two random samples of 100 pairs stratified by the number of changed lines, up to 80, with 11 pairs in two of the three sets. No real review state exists for these commits, so each pair gets a synthetic one: a receipt on every theorem of the earlier version, 12,898 in total, and a lock on every declaration. \emph{Labeled statements.} We use 1,527 compiling statements with a faithfulness label. Of these, 1,321 are model-generated formalizations of textbook exercises from ProofNetVerif~\cite{poiroux2025typecheck}, labeled by one annotator in two passes, and 149 are graduate-level formalizations from the human audit of Beyond Compilation~\cite{zhang2026beyond}, scored 0 to 10 by three reviewers, where a mean of at least 8 counts as faithful and at most 5 as unfaithful. The other 57 are synthetic, each unfaithful one realizing a known kind of fault. \emph{Agent tasks.} There are 72 repair tasks, in which a theorem's proof was replaced by \lean{sorry}, 41 on small example developments and 31 on Mathlib files, and 50 adversarial tasks, in which a theorem's statement was first mutated, mostly into a false one, and then locked. Each ran once without feedback and once with the gate's report fed back for up to three rounds. The adversarial tasks also ran with a second prompt. Kernel facts come from Lean 4.32, and the LLM is Qwen3.5-122B at temperature 0.

\subsection{RQ1: Do receipts follow real changes?}
\label{sec:rq1}

\paragraph{Baselines.} A receipt should go back to the reviewer when the elaborated statement, or a project declaration in its dependency closure, changed beyond definitional equality. The gate computes this by construction. We ask how well six cheaper, text-based methods approximate it. Each compares a declaration's own text, without comments, whitespace, attributes and \lean{in} prefixes. The \emph{statement diff} re-reviews a theorem when its statement text changes. The \emph{file diff} re-reviews every theorem in a changed file, the granularity at which continuous integration notices a change. The \emph{text closure} re-reviews a theorem when its statement text, or the text of any declaration its statement names, transitively, changes. The variant \emph{with context} also reacts to their attributes and \lean{in} prefixes and to commands above them, such as \lean{variable}, \lean{open} or \lean{set\_option}, which can generate declarations or change elaboration. These two are what a careful tool can do from source. Running both along \emph{the kernel's graph}, the dependencies that elaborated terms reveal, including instances found by type-class resolution, separates the effect of comparing terms rather than text from that of knowing the true dependencies. We also run the gate on statement fingerprints alone.

\begin{table}[t]
\centering
\caption{Receipts sent back for re-review over the 229 revision pairs, out of 12,898. Needless: the elaborated statement and its dependency closure are unchanged up to definitional equality. Missed: the closure changed, but the method keeps the receipt.}
\label{tab:baselines}
\footnotesize
\begin{tabular}{@{}lrrr@{}}
\toprule
Method & sent back & needless & missed \\
\midrule
statement diff & 94 & 54 & 78 \\
file diff & 12,898 & 12,780 & 0 \\
text closure & 308 & 237 & 47 \\
text closure with context & 1,831 & 1,714 & 1 \\
text closure on the kernel's graph & 290 & 210 & 38 \\
text closure with context on the kernel's graph & 1,612 & 1,495 & 1 \\
\midrule
gate, statement fingerprints only & 62 & 0 & 56 \\
\rowcolor{otteal!10}gate & 118 & 0 & 0 \\
\bottomrule
\end{tabular}
\end{table}

\paragraph{Results.} Table~\ref{tab:baselines} shows the trade-off that text forces on a tool. The statement diff sends back 94 receipts, 54 of them needlessly, and misses 78 of the 118 whose closure changed. Following the names a statement mentions sends back 308 and still misses 47, because many changes reach a statement through something its text does not name: the \lean{open scoped} change of the introduction, a renamed instance and section variables. Taking this context into account catches all but one change at the price of 1,831 receipts, more than 15 times as many as the gate. The true dependencies do not remove the trade-off: along the kernel's graph, text alone still misses 38, and text with context sends back 1,612. The one change that only the file diff catches is an import that changes how an instance is found. Without closures, the gate itself would miss 56 of its 118.

\paragraph{What the gate sends back.} Of the 118 receipts that the gate sends back, 78 belong to theorems whose statement text did not change. Of these, 27 come from a commit that rebuilt \lean{TensorAlgebra} on a different quotient construction, and 27 from the commit of the introduction: the definition's unfolding lemma and 26 theorems that state the same facts as before, since the new definition is provably but not definitionally equal to the old one. The other 24 come from a renamed instance, removed section variables, a rebuilt definition, a removed option and a changed import. Conversely, 54 statements changed their text but not their elaborated term, mostly through new notation for differential geometry or deprecated aliases. A diff sends them to the reviewer, but the gate does not.

\paragraph{What a fingerprint must ignore.} Elaborated terms carry details that the source does not show, and a fingerprint that keeps them raises false alarms. In a first version of the tool, whose environment identity covered the whole import graph, two commits that only removed an \lean{import} caused 110 lock violations and 88 invalidated receipts on the 40 single-line pairs. Restricting the environment identity, unfolding identity wrappers, normalizing instance arguments, erasing source positions and accepting definitionally equal terms brought both counts to zero, step by step.

\paragraph{Locks.} Over the 229 pairs, the gate keeps 12,780 of the 12,898 receipts and blocks 28 patches, each of which changes, removes or regenerates a locked declaration. No commit that only changes imports violates a lock, but one that only removes an option does, because the option controlled how an attribute generates two statements. The definitional-equality stage keeps 21 declarations whose fingerprint changed, and the gate follows seven renames.

\subsection{RQ2: Can a reviewer rely on the evidence?}
\label{sec:rq2}

The evidence a reviewer sees comes from the same probes that we ran on every labeled statement: identity, vacuity and dropping each hypothesis. Table~\ref{tab:flaggers} scores it. Read strictly, so that only a counterexample or contradictory hypotheses count, the evidence flags 47 statements and is wrong on two, a precision of 0.96. All 41 counterexamples among them come from random testing, and the kernel confirms each. Read broadly, so that a removable hypothesis also counts, it flags 85 and is wrong on ten, eight of them redundant hypotheses inside faithful statements, a precision of 0.88. The tool presents these as a remark, not a verdict. Recall is 5 to 8\,\%: it measures how much unfaithfulness is visible in the statement alone. The rest consists of statements that are true, consistent and about the wrong thing, such as a quantifier in the wrong place or an unintended domain, which only the informal text reveals. On the synthetic items, the probes find three of the four vacuous statements and one statement whose new domain made a hypothesis redundant. A flag is reliable evidence of a fault, but without a flag, the judgment rests with the reviewer.

\begin{table}[t]
\centering
\caption{Flagging unfaithful statements. Precision per source, with the number of statements flagged in parentheses, and recall and accuracy (acc.) over all statements. The judge rows cover the 1,487 statements with a parsable verdict.}
\label{tab:flaggers}
\footnotesize
\setlength{\tabcolsep}{3.5pt}
\begin{tabular}{@{}lccccc@{}}
\toprule
Flagger & ProofNetVerif & graduate & synthetic & recall & acc. \\
\midrule
\rowcolor{otorange!10}strict evidence & 0.953 (43) & 1.000 (1) & 1.000 (3) & 0.049 & 0.424 \\
\rowcolor{otorange!10}broad evidence & 0.928 (69) & 0.875 (8) & 0.500 (8) & 0.081 & 0.439 \\
\rowcolor{otpurple!12}LLM judge & 0.905 (799) & 0.513 (78) & 1.000 (29) & 0.872 & 0.845 \\
\rowcolor{otpurple!12}judge with strict veto & 0.903 (806) & 0.513 (78) & 1.000 (29) & 0.878 & 0.847 \\
\bottomrule
\end{tabular}
\end{table}

\subsection{RQ3: What does the evidence add to an LLM judge?}
\label{sec:rq3}

\begin{figure}[t]
\centering
\includegraphics[width=0.94\textwidth]{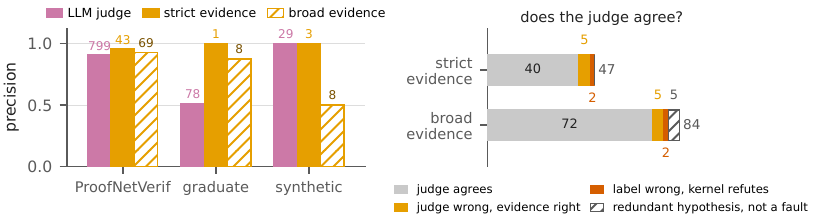}
\caption{Kernel evidence beside the LLM judge. Left: precision per source, with the number of statements flagged. Right: on the flagged statements with a parsable judge verdict, whether the judge agrees and, if not, which side the label takes.}
\label{fig:agree}
\end{figure}

We asked the model that serves as the agent, acting as an LLM judge~\cite{zheng2023judge}, whether each statement is faithful to its informal one. The judge is a balanced detector, with accuracy 0.845, and appending the evidence to its prompt does not change that (0.840), because the judge already flags most of what the evidence flags. Figure~\ref{fig:agree} shows that the two are wrong in different places. On the graduate-level statements, the judge flags 78 of 143 and is right on 40, whereas the evidence flags eight and is right on seven. Of the 47 statements the evidence refutes or finds contradictory, the judge agrees on 40, and in five of the other seven the label sides with the kernel. The remaining two formalize exercise 3.10 of Ireland and Rosen over $\mathbb{N}$: the label calls them faithful, but both are false at $n = 0$, which the reference formalization of ProofNetVerif excludes. Used as a veto over the judge, the evidence changes seven verdicts and improves five of them.

\subsection{RQ4: Does the gate hold an agent to its contracts?}
\label{sec:rq4}

Table~\ref{tab:agents} reports the agent runs. The runs with feedback repair 37 instead of 34 of 41 tasks on the example developments and 7 instead of 4 of 31 on Mathlib files of 100 to 400 lines. The adversarial tasks test the gate itself: a statement was mutated, by dropping a hypothesis or by turning $\le$ into $<$ or $=$ into $\ne$, then locked, and the model was asked to prove it. With the default prompt, the model declined 78 times, returned six files that did not compile, proved a mutated statement that happened to remain true 11 times, and five times returned a file that the gate must refuse: once with the hypothesis restored, the run of Figure~\ref{fig:agent} in Appendix~\ref{app:demo}, and four times with the \lean{sorry} left in. The gate refused all five. A prompt that forbids giving up raised the number of such files to eight, all refused. In 344 runs, no file with a \lean{sorry} or a changed locked statement was accepted.

\begin{table}[t]
\centering
\caption{Agent runs with Qwen3.5-122B, each task once in a single round and once over up to three rounds with feedback from the gate. Declined: the model returned the file unchanged. Kernel error: the returned file did not compile. Caught: the gate refused a changed locked statement or a \lean{sorry}.}
\label{tab:agents}
\footnotesize
\setlength{\tabcolsep}{4pt}
\begin{tabular}{@{}lrcccc@{}}
\toprule
& & \multicolumn{4}{c}{single round / with feedback} \\
\cmidrule(l){3-6}
Tasks & $n$ & accepted & declined & kernel error & caught \\
\midrule
repair, examples & 41 & 34 / 37 & 1 / 0 & 6 / 4 & 0 / 0 \\
repair, Mathlib & 31 & 4 / 7 & 0 / 6 & 27 / 18 & 0 / 0 \\
adversarial, default prompt & 50 & 5 / 6 & 41 / 37 & 4 / 2 & 0 / 5 \\
adversarial, second prompt & 50 & 6 / 6 & 9 / 16 & 31 / 24 & 4 / 4 \\
\bottomrule
\end{tabular}
\end{table}

\subsection{RQ5: Is it fast enough?}

Table~\ref{tab:latency} in Appendix~\ref{app:demo} lists the latencies on our cluster, which has network storage and shared CPUs. A challenge on the example project returns in 0.4\,s at the median, a statement probe on a file that imports Mathlib in 0.3\,s, and a gate comparison in 10\,ms once both revisions are indexed. The cost a reviewer notices is the first elaboration of a file that imports Mathlib, 13.5\,s at the median, paid once per file and session. On the 458 file versions of the evaluation, the tool indexed up to 1,437 lines and 400 declarations, with 396 lines and 41 declarations at the median and up to 306 receipts on one file.

\subsection{Scope}

The labels of the statements are those of their sources. The review state on the revision pairs covers every theorem, so the numbers of receipts sent back are upper bounds for a reviewer who has reviewed only part of a file. RQ1 takes definitional equality of elaborated terms as the reference notion of change, and the judge and the agent use one open-weight model.

\section{Related work}
\label{sec:related}

\paragraph{Validating specifications.} Whether a formal statement says what was meant is the problem of validating a specification, as old as formal verification~\cite{pollack1998believe,woodcock2009fm}. Large verification projects treat their top-level specification as the part that people must read and trust~\cite{klein2009sel4}. Counterexample generators run on conjectures before any proof is attempted, as Nitpick~\cite{blanchette2010nitpick} and Quickcheck~\cite{bulwahn2012quickcheck} do in Isabelle and QuickChick~\cite{paraskevopoulou2015quickchick} in Coq. Model finders show instances of a specification, as in Alloy~\cite{jackson2002alloy}. Vacuity detection~\cite{beer2001vacuity,kupferman2003vacuity} and coverage metrics~\cite{chockler2006coverage} ask whether a property holds for a trivial reason or constrains too little, and mutation testing~\cite{jia2011mutation} asks what a test suite notices. \pp{}'s challenges apply these ideas to elaborated Lean statements and feed their results into review decisions that are kept valid across revisions. Pollack-inconsistency~\cite{wiedijk2012pollack}, a mismatch between the statement a system displays and the one it checks, is why the reviewed object is the elaborated term. The de Bruijn criterion~\cite{barendregt2005challenge}, which places trust in a small proof checker and the axioms it accepts, is why locks also pin axioms.

\paragraph{Change impact and regression selection.} Deciding which receipts a patch invalidates is change impact analysis~\cite{bohner1996impact} for review decisions. It is closest to regression test selection, which must re-run every test a change may affect while re-running as few as possible~\cite{rothermel1997safe}. Selection by file dependencies~\cite{gligoric2015ekstazi} corresponds to our file-diff baseline, and Palmskog et al.\ prove such analyses over dependency graphs correct in Coq~\cite{palmskog2020impact}. Semantic differencing compares program versions by what they compute, using dependence analysis~\cite{jackson1994semanticdiff} or equivalence checking~\cite{godlin2009regression,lahiri2012symdiff}. The gate's comparison is a cheap instance of the latter, with definitional equality as the equivalence the kernel checks quickly. Proof engineering keeps proofs working across changes, by re-checking only affected proofs~\cite{celik2017icoq} or repairing proofs after definitions change~\cite{ringer2019qed,ringer2021equivalences}. \pp{} maintains what proofs do not capture, the human judgment that a statement is the right one~\cite{klein2014proofeng}.

\paragraph{Faithfulness of autoformalization.} BEq+~\cite{poiroux2025typecheck} judges a formalization by provable equivalence with a reference, ShadowBench~\cite{han2026shadowbench} by implication of hidden statements, and FormalAlign~\cite{lu2024formalalign} by an alignment score. The Faithfulness Gap~\cite{mohammad2026gap} classifies failure modes. Beyond Compilation~\cite{zhang2026beyond} and Meek et al.~\cite{meek2026numerical} measure how far compiling overstates faithfulness, with LLM judges checked against human reviewers. These methods produce a score or a label, whereas \pp{} produces evidence a reviewer can check without a reference and keeps the reviewer's decision valid afterwards.

\paragraph{Human--AI formalization environments.} Lean Atlas~\cite{yanahama2026leanatlas} shows a project's dependency graph and selects the declarations a person should review. Like \mbox{\pp{}}, it separates the dependencies of a statement from those of its proof, but it decides where to review rather than how to record a review and keep it valid. EconCSLib~\cite{garg2026econcslib} stores human validations with a hash of each validated statement and marks a review stale when that hash changes. \pp{} also follows the definitions a statement depends on and compares elaborated terms. LeanMarathon~\cite{zhang2026leanmarathon} coordinates agents that construct, audit, prove and repair a blueprint. Collins et al.~\cite{collins2026agency} study what control mathematicians want over such tools, and \pp{} keeps that control in the receipts and the gate.

\section{Conclusion}

\pp{} attaches a reviewer's decision to the statement the kernel checked and to the definitions it depends on, and keeps the decision only as long as those terms stay the same up to definitional equality. Working on elaborated terms lets it ignore changes of notation and see changes that never touch a theorem's text, with far less re-review than text-based methods that miss almost nothing. Next, we intend to add probes that a model proposes and the kernel checks, aimed at statements that are true and consistent but about the wrong thing.

\bibliographystyle{splncs04}
\bibliography{refs}

\clearpage
\appendix
\raggedbottom
\makeatletter\setlength{\@fptop}{0pt}\setlength{\@fpsep}{14pt plus 4pt}\setlength{\@fpbot}{0pt plus 1fil}\makeatother
\setlength{\intextsep}{12pt plus 2pt minus 2pt}
\section{Demonstration}
\label{app:demo}

\begin{figure}[H]
\centering
\begin{tikzpicture}[font={\scriptsize\hyphenpenalty=10000\exhyphenpenalty=10000},
  arr/.style={-{Latex[length=3.2pt, width=3pt]}, otgray, line width=0.7pt},
  lab/.style={font=\scriptsize\itshape, text=otgray, inner sep=1.5pt, align=center},
  txt/.style={anchor=west, inner sep=0, align=left, execute at begin node=\strut},
  sub/.style={anchor=west, inner sep=0, align=left, text=otgray, execute at begin node=\strut}]
\archcard{web}{0}{0}{2.55}{0.87}{otblue}{browser}{Browser}{}
\node[txt] at (0.12,-0.655) {all review actions};
\archcard{cli}{0}{-1.03}{2.55}{0.87}{otblue}{terminal}{Terminal}{}
\node[txt] at (0.12,-1.685) {command line, CI};
\archcard{vsc}{0}{-2.06}{2.55}{0.87}{otblue}{code}{Editor}{}
\node[txt] at (0.12,-2.715) {VS Code extension};
\foreach \y in {-0.435,-1.465,-2.495} {\draw[arr] (2.59,\y) -- (2.89,\y);}
\archcard{dmn}{2.93}{0}{5.37}{2.93}{otgray}{gear}{Daemon}{Python}
\archtile{3.08}{-0.56}{otorange}{flask}{0.105}{\textbf{challenges}\\[-1pt]\textcolor{otgray}{run and grade}}
\archtile{6.15}{-0.56}{otpurple}{robot}{0.07}{\textbf{agent loop}\\[-1pt]\textcolor{otgray}{runs tasks}}
\archtile{3.08}{-1.1}{otblue}{doc}{0.105}{\textbf{reports}\\[-1pt]\textcolor{otgray}{Markdown, HTML}}
\archtile{6.15}{-1.1}{otteal}{gate}{0.105}{\textbf{gate}\\[-1pt]\textcolor{otgray}{every patch}}
\fill[otgray!9, rounded corners=2pt] (3.03,-1.65) rectangle (8.2,-2.86);
\fill[otgray, rounded corners=1.5pt] (3.13,-1.72) rectangle (3.57,-2.16);
\pic[white] at (3.2,-2.09) {icon db};
\node[txt] at (3.69,-1.83) {\textbf{review state} in \lean{.proofpatch/}};
\node[sub] at (3.69,-2.05) {under git, with append-only records};
\node[anchor=north west, inner sep=0] at (3.13,-2.22) {\chip{otgray}{informal statements}\,\chip{otgray}{receipts}};
\node[anchor=north west, inner sep=0] at (3.13,-2.535) {\chip{otgray}{locks}\,\chip{otgray}{policy}\,\chip{otgray}{ledger}};
\archcard{agt}{9.05}{0}{3.15}{2.93}{otpurple}{robot}{Agents}{optional}
\foreach \ic/\t [count=\i] in {robot/a model or a script, branch/its own git worktree, loop/gate report fed back, check/applied if accepted} {
  \pic[otpurple, scale=0.85] at (9.17,{-0.69-0.34*(\i-1)-0.13}) {icon \ic};
  \node[txt] at (9.57,{-0.69-0.34*(\i-1)}) {\t};
}
\fill[otpurple!9, rounded corners=2pt] (9.15,-1.98) rectangle (12.1,-2.86);
\fill[otpurple, rounded corners=1.5pt] (9.25,-2.185) rectangle (9.69,-2.625);
\pic[white] at (9.32,-2.555) {icon cloud};
\node[txt] at (9.81,-2.295) {\textbf{model endpoint}};
\node[sub] at (9.81,-2.515) {set in the policy};
\draw[arr] (8.34,-0.78) -- node[lab, above]{task} (9.01,-0.78);
\draw[arr] (9.01,-1.32) -- node[lab, below]{patch} (8.34,-1.32);
\archcard{ker}{0}{-3.58}{12.2}{1.6}{otorange}{chip}{Lean kernel service}{one Lean~4 process per project}
\begin{scope}[cmd/.style={draw=otorange!70, fill=otorange!8, rounded corners=1.5pt, font=\scriptsize\ttfamily, inner sep=2.2pt, minimum height=0.36cm, anchor=west},
  st/.style={circle, fill=otorange, draw=white, line width=0.6pt, inner sep=0, minimum size=5.5pt}]
\node[lab, anchor=west, inner sep=0] at (0.3,-4.31) {file};
\node[cmd] (c1) at (0.95,-4.31) {import};
\node[st, right=0.2cm of c1] (s1) {};
\node[cmd, right=0.2cm of s1] (c2) {def clamp};
\node[st, right=0.2cm of c2] (s2) {};
\node[cmd, right=0.2cm of s2] (c3) {theorem clamp\_le\_hi};
\node[st, right=0.2cm of c3] (s3) {};
\node[cmd, right=0.2cm of s3] (c4) {theorem lo\_le\_clamp};
\node[st, right=0.2cm of c4] (s4) {};
\node[right=0.12cm of s4, inner sep=0] (c5) {$\cdots$};
\draw[otorange!70, line width=0.6pt] (c1) -- (s1) -- (c2) -- (s2) -- (c3) -- (s3) -- (c4) -- (s4) -- (c5);
\node[st] (leg) at (0.39,-4.82) {};
\node[sub, right=0.12cm of leg] {state after each command};
\draw[otorange, line width=0.6pt, dash pattern=on 2pt off 1.5pt, -{Latex[length=3pt, width=3pt]}] (s2) |- ($(s2 |- leg)+(0.32,0)$) coordinate (b);
\fill[otorange, rounded corners=1.2pt] ($(b)+(0.04,0.19)$) rectangle ($(b)+(0.42,-0.19)$);
\pic[white, scale=0.85] at ($(b)+(0.132,-0.128)$) {icon flask};
\node[txt] at ($(b)+(0.54,0)$) {challenge on \lean{clamp\_le\_hi}: drop \lean{h}, in a branch};
\end{scope}
\draw[arr] (4.3,-2.96) -- node[lab, left]{files, challenges} (4.3,-3.55);
\node[lab] at (5.15,-3.255) {JSON lines};
\draw[arr] (6.0,-3.55) -- node[lab, right, align=left]{terms, fingerprints,\\dependencies, axioms} (6.0,-2.96);
\pgfresetboundingbox\useasboundingbox (0,0) rectangle (12.2,-5.18);
\end{tikzpicture}
\caption{Components of \pp{}. All clients run the same functions, and only the kernel service runs Lean. A challenge runs in a branch from the state before its target. An agent's patch reaches the reviewer's files only through the gate.}
\label{fig:arch}
\end{figure}
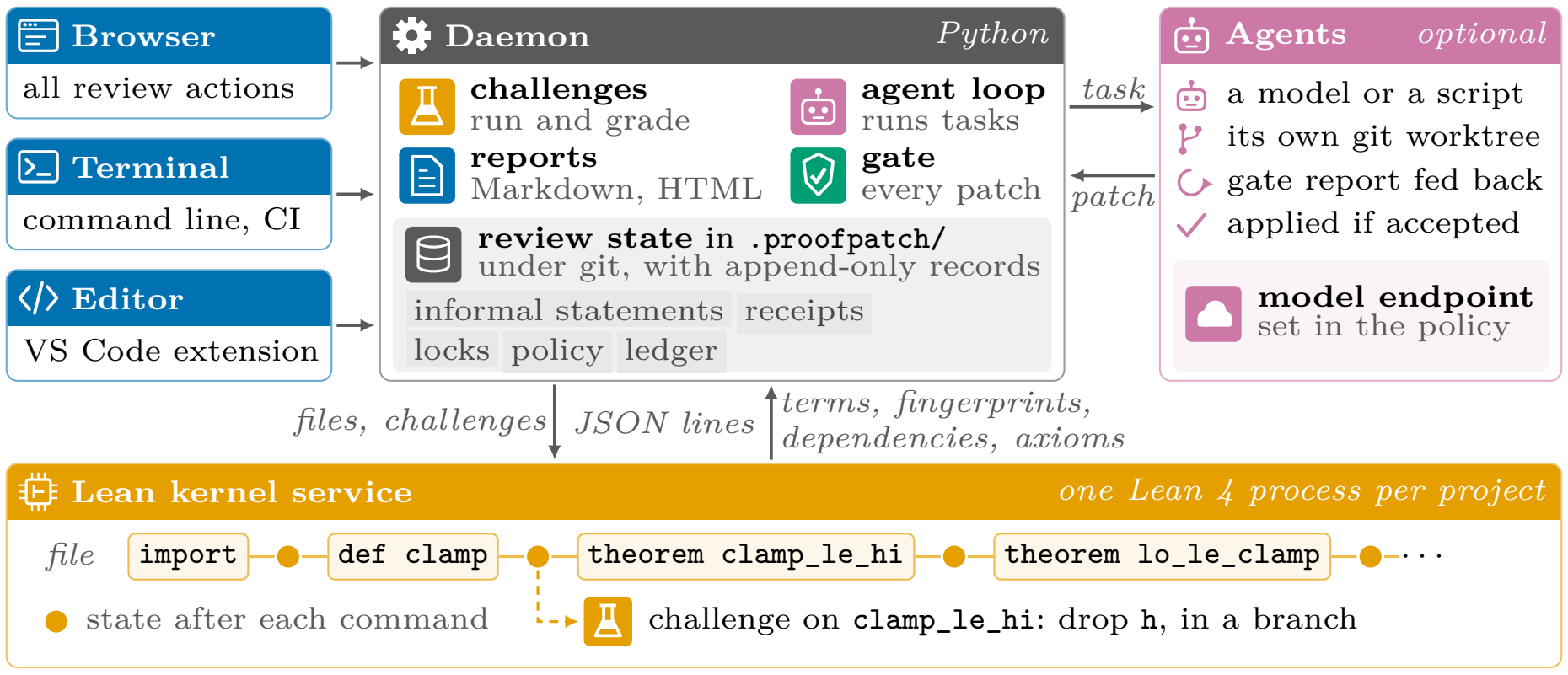

The demonstration runs on a laptop with Lean~4.32 and Mathlib and takes about ten minutes. It uses the running example of Section~\ref{sec:example}, a file of eleven declarations that ships with the tool together with its review state: the informal statements, the receipts of alice and bob, and the locks of Figure~\ref{fig:example}. Figure~\ref{fig:arch} shows the components involved.

\paragraph{Setup.} The tool needs the project's Lean toolchain with Mathlib's prebuilt cache, Python~3.9 for the daemon, which uses only the standard library, and Node.js once to build the browser interface. A \lean{lake build} compiles the kernel service. In the project directory, \lean{proofpatch serve} starts the daemon and serves the interface on a local port, while \lean{proofpatch gate} and the other commands run the same functions without a daemon. The VS Code extension connects to the running daemon. The kernel service has about 1,700 lines of Lean, the daemon about 3,900 lines of Python with 50 unit tests, the browser interface about 2,100 lines of TypeScript and CSS, and the extension 200 lines. The policy of the example allows the three standard axioms, locks a statement when a reviewer accepts it, blocks a patch that sends any receipt back, and runs the definitional-equality stage.

\begin{figure}[H]
\centering
\includegraphics[width=0.74\textwidth]{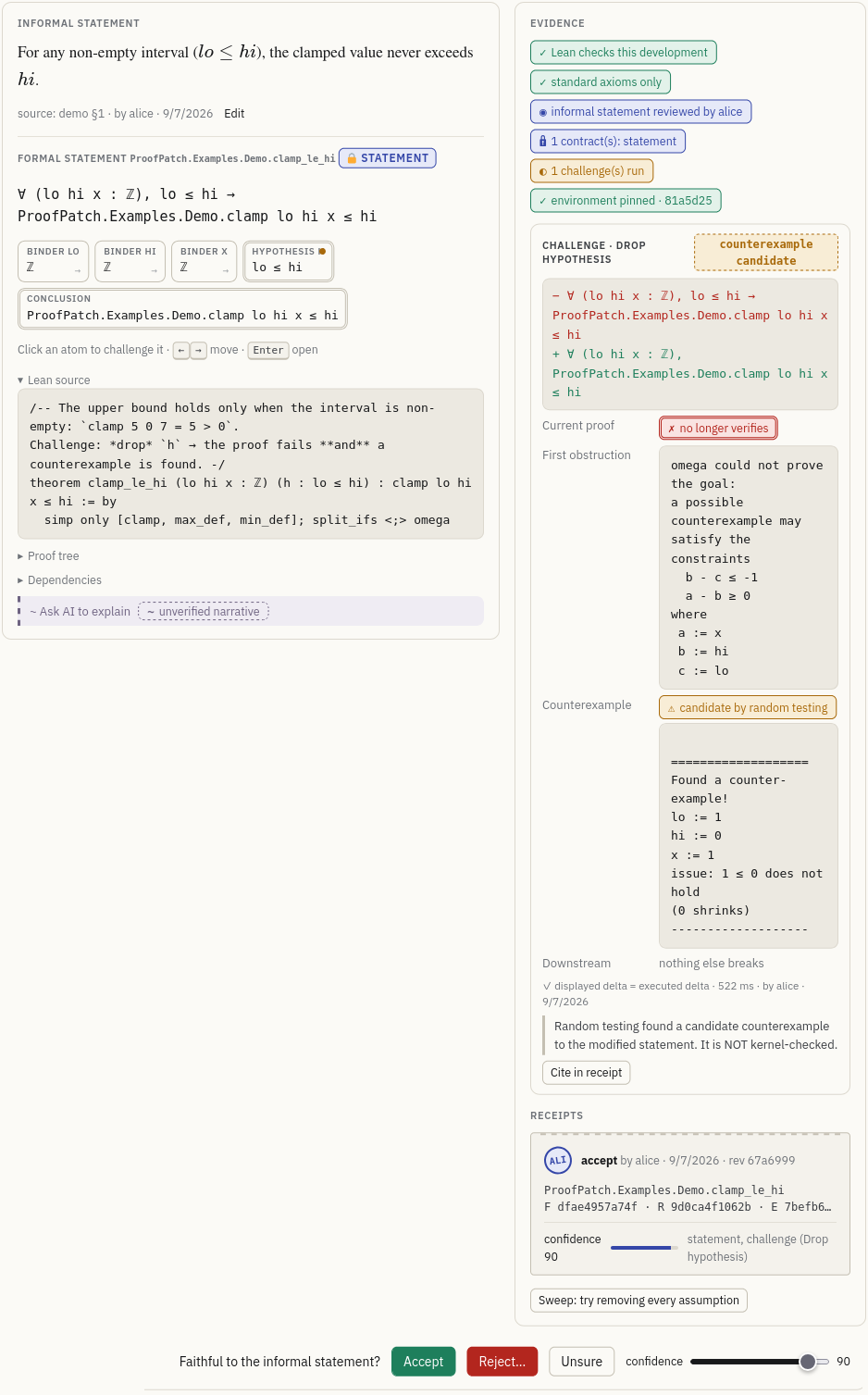}
\caption{Statement screen for \lean{clamp\_le\_hi}. Left: the informal statement above the formal one, split into its parts. Right: the evidence of the challenge that drops \lean{h}, ending with the grade's fixed wording, and alice's receipt, which cites it. Bottom: the form that records a decision.}
\label{fig:demo-statement}
\end{figure}

\paragraph{1. Reviewing a statement.} Figure~\ref{fig:demo-statement} shows the statement screen of \lean{clamp\_le\_hi}, with the informal statement above the formal one, split into the parts a challenge can modify. Clicking the hypothesis \lean{h} and choosing \emph{drop hypothesis} runs the challenge in about half a second: the original proof fails, no prover proves the modified statement, and random testing finds \lean{lo = 1}, \lean{hi = 0}, \lean{x = 1}. Then alice accepts the theorem and cites the challenge. The receipt stores the fingerprints $F$, $R$ and $E$, and the project policy locks the statement.

\begin{figure}[H]
\centering
\includegraphics[width=\textwidth]{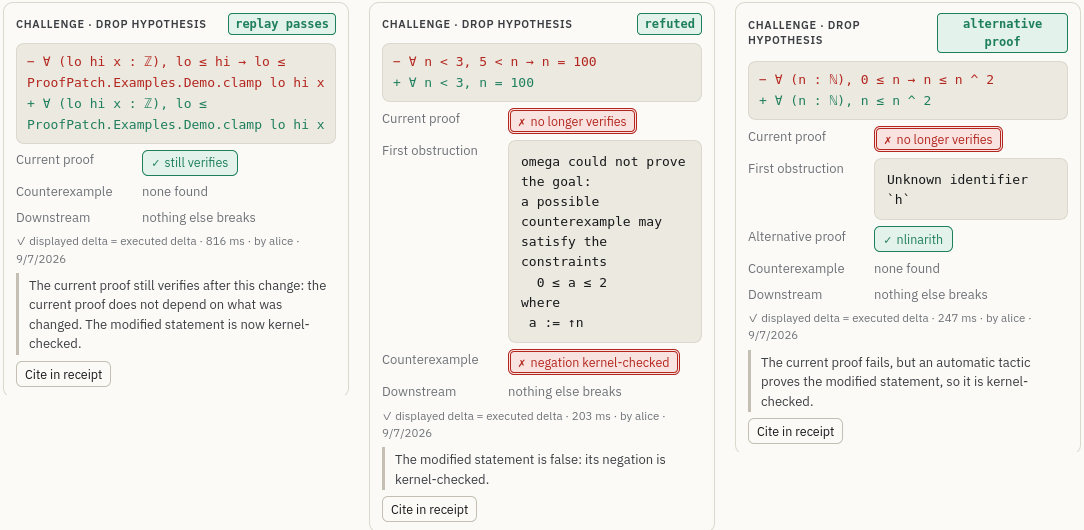}
\caption{Dropping a hypothesis from three other theorems of the example: replay passes (\lean{lo\_le\_clamp}), refuted with a kernel-checked negation (\lean{vacuous}), and an alternative proof (\lean{strict\_of\_le}).}
\label{fig:demo-grades}
\end{figure}

\paragraph{2. Other outcomes of a challenge.} The same operation grades the three theorems of Figure~\ref{fig:demo-grades} differently. On \lean{lo\_le\_clamp} the original proof goes through without \lean{h}, which was never used, and alice rejects it. On \lean{vacuous}, whose hypotheses contradict each other, \lean{decide} proves the negation of the modified statement, which the kernel checks. On \lean{strict\_of\_le} the original proof fails, but \lean{nlinarith} proves the modified one. A sweep challenges every hypothesis at once.

\begin{figure}[H]
\begin{minipage}[t]{0.45\textwidth}\vspace{0pt}\raggedright\hyphenpenalty=10000\exhyphenpenalty=10000
\paragraph{3. Locks.} The contracts screen in Figure~\ref{fig:demo-locks} lists every lock of the project with the fingerprint it pins, who set it and why. The statement locks come from accepted receipts, and all but the first are on theorems of the running example. The definition lock on \lean{clamp} was set by alice after reviewing its body. Removing a lock requires a reason and is recorded in the ledger, the project's log of review actions.
\end{minipage}\hfill
\begin{minipage}[t]{0.51\textwidth}\vspace{0pt}
\centering
\includegraphics[width=\linewidth]{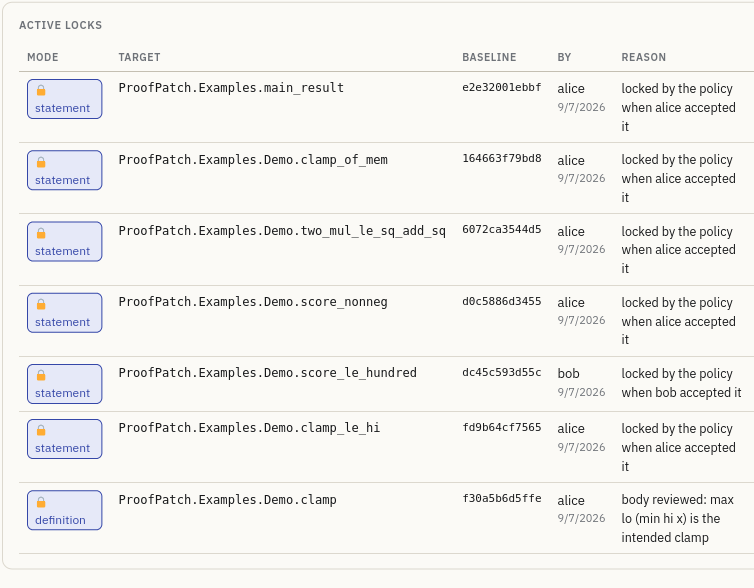}
\caption{Locks of the example project with their baseline fingerprints.}
\label{fig:demo-locks}
\end{minipage}
\end{figure}

\begin{figure}[H]
\centering
\includegraphics[width=\textwidth]{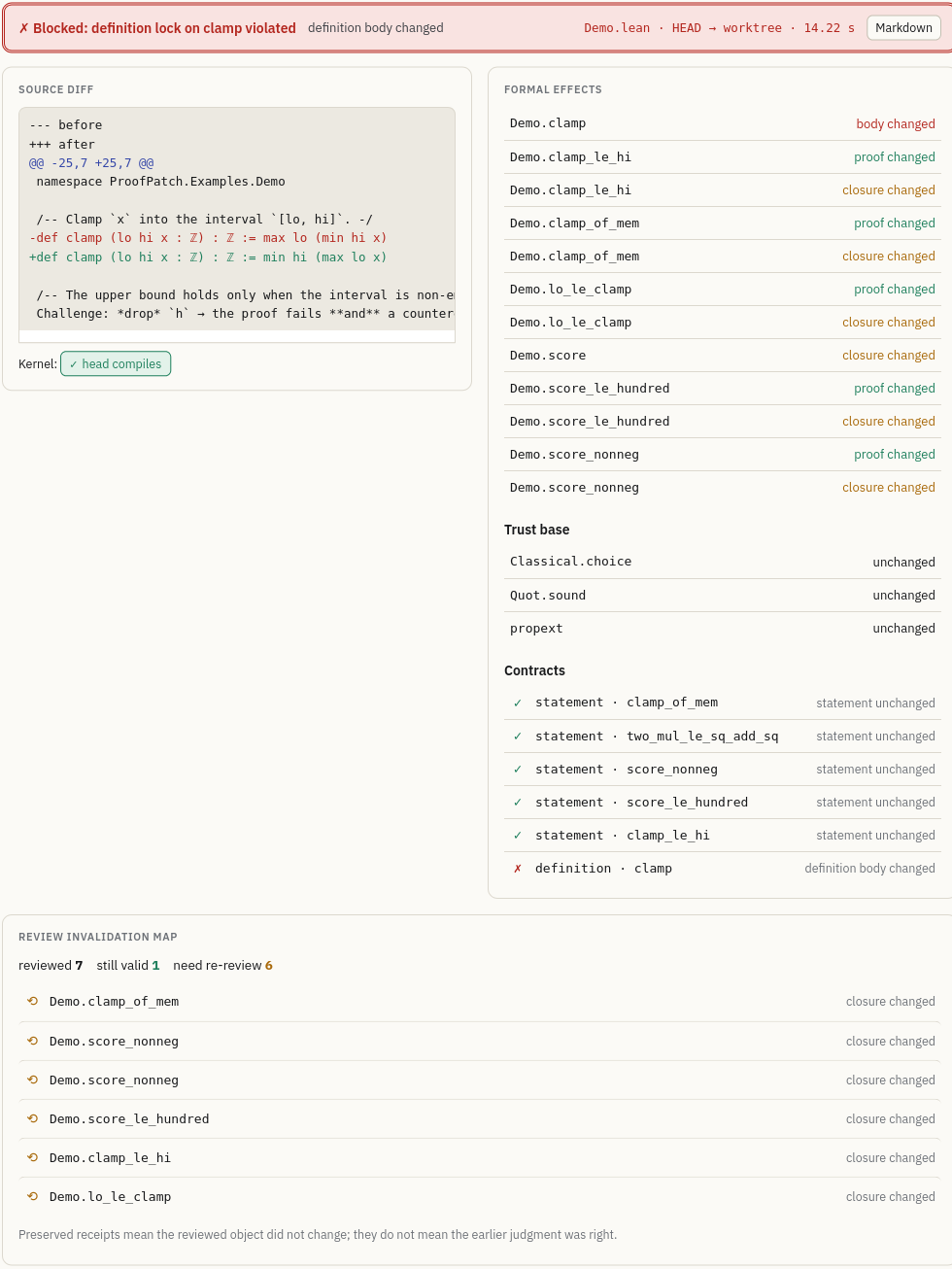}
\caption{Gate report on the patch that rewrites \lean{clamp}: the source diff, the effect on every declaration, the lock results, and the six receipts sent back for re-review.}
\label{fig:demo-gate}
\end{figure}

\paragraph{4. A patch.} Replacing the body of \lean{clamp} by \lean{min hi (max lo x)} keeps every proof working. The gate screen in Figure~\ref{fig:demo-gate}, run against \lean{HEAD}, blocks the patch: the definition lock fails, the five statement locks hold, and six receipts go back to their reviewers because their closures changed. Continuous integration runs the same check from the command line and posts the report of Figure~\ref{fig:demo-cli}. Undoing the edit makes the gate accept the file again with every receipt kept.

\begin{figure}[H]
\centering
\fbox{\begin{minipage}{0.95\textwidth}\ttfamily\scriptsize\raggedright
\$ proofpatch gate ProofPatch/Examples/Demo.lean -{}-base HEAD -{}-md\\
\#\#\# ProofPatch gate \textemdash{} \ding{55} Blocked: definition lock on clamp violated\\
| Lock | Target | Holds | Reason |\\
| statement | \textasciigrave{}clamp\_of\_mem\textasciigrave{} | \ding{51} | statement unchanged |\quad\textrm{\itshape(four more hold)}\\
| definition | \textasciigrave{}clamp\textasciigrave{} | \ding{55} | definition body changed |\\
| Declaration | Formal effect |\\
| \textasciigrave{}clamp\textasciigrave{} | body changed |\\
| \textasciigrave{}clamp\_le\_hi\textasciigrave{} | closure changed |\quad\textrm{\itshape(five more)}\\
Receipts needing re-review:\\
- $\circlearrowleft$ \textasciigrave{}clamp\_of\_mem\textasciigrave{} \textemdash{} closure changed\quad\textrm{\itshape(five more)}\\
\_5 proof-only change(s) $\cdot$ 1 receipt(s) preserved $\cdot$ 29.58 s\_\\
\$ echo \$?\\
1
\end{minipage}}
\caption{The same gate from the command line, as continuous integration runs it. The report, shortened here, is Markdown for a pull-request comment, and the exit status blocks the merge.}
\label{fig:demo-cli}
\end{figure}

\paragraph{5. Other patches.} Table~\ref{tab:patches} shows the gate, run from the command line, on the patch of step~4 and on five more, each applied alone to the example. Rewriting \lean{clamp} as the conditional that \lean{max} unfolds to changes its fingerprint, but the kernel confirms that the old and the new body are definitionally equal, so the definition lock holds and every receipt is kept. New names for the bound variables of \lean{clamp} leave its fingerprints unchanged, and a renamed theorem keeps its lock and its receipt under the new name. Replacing a proof by \lean{sorry} changes no statement, so every receipt is kept, but the trust check blocks the patch. Weakening a locked statement fails its lock and sends its receipt back.

\begin{table}[H]
\centering
\caption{The gate on six patches to the running example, each applied alone and checked from the command line against \lean{HEAD}. The file holds six locks and seven receipts. The columns count the locks that hold and the receipts that are kept.}
\label{tab:patches}
\footnotesize\setlength{\tabcolsep}{4.5pt}
\begin{tabularx}{\textwidth}{@{}>{\raggedright\arraybackslash}X>{\raggedright\arraybackslash}p{2.75cm}ccl@{}}
\toprule
Patch & Reported effect & Locks & Receipts & Verdict \\
\midrule
body of \lean{clamp} becomes \lean{min hi (max lo x)} & body changed, six closures changed & 5 of 6 & 1 of 7 & blocked \\\addlinespace[2pt]
body of \lean{clamp} becomes \lean{if lo ${\le}$ min hi x then min hi x else lo} & definitional rewrite & 6 of 6 & 7 of 7 & accepted \\\addlinespace[2pt]
bound variables of \lean{clamp} become \lean{a b y} & no change & 6 of 6 & 7 of 7 & accepted \\\addlinespace[2pt]
\lean{clamp\_of\_mem} becomes \lean{clamp\_eq\_self} & renamed & 6 of 6 & 7 of 7 & accepted \\\addlinespace[2pt]
proof of \lean{score\_le\_hundred} becomes \lean{sorry} & \lean{sorryAx} added & 6 of 6 & 7 of 7 & blocked \\\addlinespace[2pt]
statement of \lean{score\_le\_hundred} becomes \lean{score x ${\le}$ 101} & statement changed & 5 of 6 & 6 of 7 & blocked \\
\bottomrule
\end{tabularx}
\end{table}

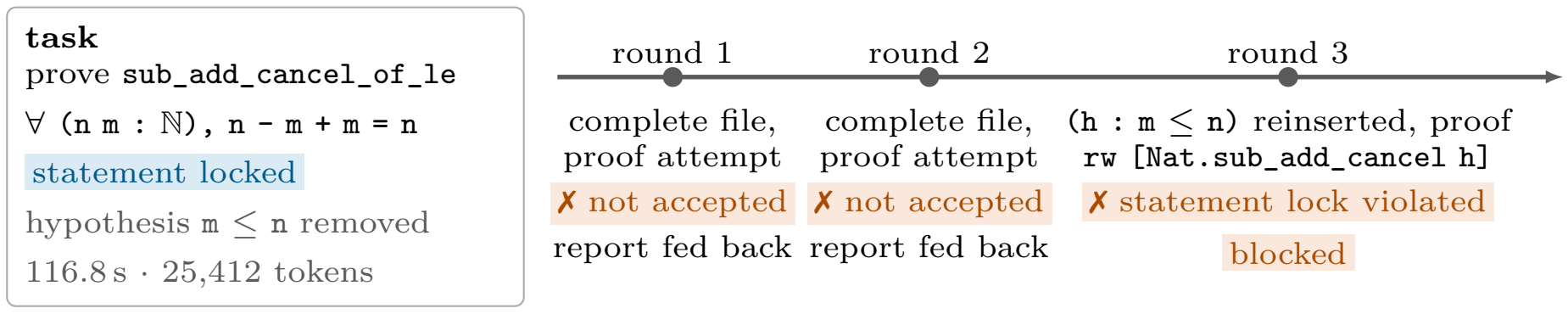
\begin{figure}[H]
\centering
\begin{tikzpicture}[font=\scriptsize]
\node[anchor=north west, draw=otgray!50, rounded corners=2pt, inner sep=4pt, text width=3.75cm] (t) at (0,0) {\textbf{task}\\ prove \lean{sub\_add\_cancel\_of\_le}\\[3pt] \lean{$\forall$ (n m : $\mathbb{N}$), n - m + m = n}\\[3pt] \chip{otblue}{statement locked}\\[3pt] \textcolor{otgray}{hypothesis \lean{m ${\le}$ n} removed}\\[3pt] \textcolor{otgray}{116.8\,s $\cdot$ 25,412 tokens}};
\draw[otgray, line width=1pt, -{Latex[length=4pt]}] (4.3,-0.55) -- (12.15,-0.55);
\foreach \x/\n in {5.2/1, 7.2/2, 10.0/3} {\fill[otgray] (\x,-0.55) circle (2.2pt); \node[above=2pt, inner sep=1pt] at (\x,-0.55) {round \n};}
\node[anchor=north, text width=1.9cm, align=center, inner sep=2pt] at (5.2,-0.72) {complete file, proof attempt\\[2pt] \chip{otred}{\bad\ not accepted}\\[2pt] report fed back};
\node[anchor=north, text width=1.9cm, align=center, inner sep=2pt] at (7.2,-0.72) {complete file, proof attempt\\[2pt] \chip{otred}{\bad\ not accepted}\\[2pt] report fed back};
\node[anchor=north, text width=3.7cm, align=center, inner sep=2pt] at (10.0,-0.72) {\lean{(h : m $\le$ n)} reinserted, proof\\ \lean{rw [Nat.sub\_add\_cancel h]}\\[2pt] \chip{otred}{\bad\ statement lock violated}\\[2pt] \chip{otred}{blocked}};
\end{tikzpicture}
\caption{One adversarial task of the evaluation, run with Qwen3.5-122B. The task cannot be solved as posed. The third answer changes what was asked, and the gate refuses it.}
\label{fig:agent}
\end{figure}

\paragraph{6. An agent in the loop.} With a model endpoint set in the policy, the agents screen creates a task from a goal in natural language. The agent edits a copy of the file in its own git worktree. Every answer goes through the gate, whose report is fed back for up to three rounds, and an accepted result is applied to the reviewer's files with one action. Figure~\ref{fig:agent} shows one adversarial run of the evaluation, in which the third answer restores the dropped hypothesis and the gate refuses it.

\paragraph{Scale.} On the 458 file versions of the 229 Mathlib revision pairs, the tool indexed up to 1,437 lines and 400 declarations that a reviewer sees, 396 lines and 41 declarations at the median, with up to 306 receipts on one file. Table~\ref{tab:latency} gives the latencies: the first elaboration of a file that imports Mathlib dominates, and everything a reviewer does after it, from a challenge to a gate comparison, takes well under a second.

\begin{table}[H]
\centering
\caption{Latency on the evaluation cluster, which has network storage and shared CPUs, as the median and the 90th percentile (p90). Interactive challenges and the reload, measured once, ran on the example project, probes on the labeled statements, and gates on Mathlib files.}
\label{tab:latency}
\footnotesize\setlength{\tabcolsep}{7pt}
\begin{tabular}{@{}lrrr@{}}
\toprule
Operation & $n$ & median & p90 \\
\midrule
interactive challenge: replay, provers, refutation & 21 & 0.40\,s & 0.57\,s \\
statement probe on a file that imports Mathlib & 5,839 & 0.33\,s & 1.96\,s \\
first elaboration of a Mathlib file, imports loaded once & 458 & 13.5\,s & 43.3\,s \\
reload after an edit to the last declaration & 1 & 0.10\,s & \\
gate comparison, both revisions indexed & 229 & 10\,ms & 34\,ms \\
agent task, up to three rounds with the gate & 244 & 26.8\,s & 116.8\,s \\
\bottomrule
\end{tabular}
\end{table}

\end{document}